\documentclass[letterpaper, 10 pt, conference]{ieeeconf}

\IEEEoverridecommandlockouts

\usepackage{siunitx}

\usepackage[hidelinks, bookmarks=false]{hyperref}

\usepackage{graphicx}
\usepackage{mathptmx}
\usepackage{amsmath}
\usepackage{amssymb}
\usepackage{booktabs}

\usepackage{tikz}
\usetikzlibrary{arrows.meta,decorations.markings}

\newcommand{\mkcirc}[2]{\filldraw (#1,#2) circle (0.057);}
\newcommand{\mksq}[2]{\filldraw (#1-0.051,#2-0.051) rectangle (#1+0.051,#2+0.051);}
\newcommand{\mktri}[2]{\filldraw (#1,#2+0.068) -- (#1-0.062,#2-0.040)
                                  -- (#1+0.062,#2-0.040) -- cycle;}

\usepackage{amsthm}
\newtheorem{theorem}{Theorem}
\newtheorem{lemma}[theorem]{Lemma}
\newtheorem{proposition}[theorem]{Proposition}

\newtheorem{assumptionx}[theorem]{Assumption}
\theoremstyle{remark}
\newtheorem{remark}[theorem]{Remark}

\let\labelindent\relax
\usepackage{enumitem}

\newcommand{\sig}{\varphi}
\DeclareMathOperator{\sat}{sat}
\DeclareMathOperator{\sech}{sech}
\DeclareMathOperator{\Real}{Re}
\DeclareMathOperator{\artanh}{arctanh}
\DeclareMathOperator{\sign}{sign}

\title{\LARGE \bf
Types of Bursting with a Two-Block Spiking Primitive
}

\renewcommand{\authorrefmark}[1]{\raisebox{0pt}[0pt][0pt]{\textsuperscript{\footnotesize #1}}}

\author{Arthur Fyon\authorrefmark{1}, Julien Brandoit\authorrefmark{1}, Loris Mendolia\authorrefmark{1}, Pierre Sacré\authorrefmark{1}, Guillaume Drion\authorrefmark{1}, \\Alessio Franci\authorrefmark{1,2}, Christian Fernandez Lorden\authorrefmark{1}, and Gustave Bainier\authorrefmark{1}%
\thanks{This work was supported by the Belgian Government through the Federal Public Service Policy and Support (BOSA) grant NEMODEI. A.\ Fyon is a Postdoctoral Researcher of the Fonds de la Recherche Scientifique--FNRS. J. Brandoit and L. Mendolia are FRIA Grantees of the Fonds de la Recherche Scientifique--FNRS. \authorrefmark{1}A.\ Fyon, J.\ Brandoit, L.\ Mendolia, P.\ Sacré, G.\ Drion, A.\ Franci, C.\ Fernandez Lorden, and G.\ Bainier are with the Department of Electrical Engineering and Computer Science, University of Liège, 4000 Liège, Belgium. \authorrefmark{2}A.\ Franci is with the WEL-T Department, WEL Research Institute, 1300 Wavre, Belgium. \{afyon, jbrandoit, lmendolia, p.sacre, gdrion, afranci, christian.fernandezlorden, gustave.bainier\}@uliege.be.}%
}
\begin{document}

\maketitle
\thispagestyle{empty}
\pagestyle{empty}

\begin{abstract}
Excitable spikers are attractive systems for sparse, event-driven computation. A spiking primitive composed of a linear frequency selector closed in positive feedback around a high-gain saturating nonlinearity has recently been proposed as a minimal excitable system. This paper asks which bursting classes that primitive can produce with two elementary selectors: a first-order high-pass filter, and a second-order resonant band-pass filter. The nonlinearity decides whether the system fires, and where the rheobase sits. The selector decides whether rest and repetitive firing can coexist, and therefore which bursting class is available. We show that the high-pass never makes rest and firing coexist, whatever the nonlinearity, and that no linear slow feedback branch can then produce bursting. With a piecewise linear nonlinearity, it gives type~I excitability, and a second copy of the primitive gives parabolic bursting. The resonant band-pass gives a Hopf, type~II excitability, and a subcritical onset above a computable gain, hence a window where rest and firing coexist. A single additional slow linear feedback then gives elliptic bursting. We also show that the resting point is unique for both selectors, which makes square-wave bursting structurally unavailable. Closed-form thresholds, firing rates, and criticality conditions are given and validated numerically.
\end{abstract}

\section{Introduction}

Excitability is a natural spiking mechanism for sparse, event-driven computation, attractive wherever energy efficiency is a design objective~\cite{Sepulchre2022}. Rather than responding continuously, an excitable system remains quiescent until a sufficiently large perturbation triggers a brief, all-or-none excursion before it autonomously returns to rest~\cite{Sepulchre2018, Sepulchre2025}. In biological nervous systems, this mechanism underlies the generation of action potentials, the information primitive of neural computation~\cite{Hodgkin1952}, and it has inspired neuromorphic engineering~\cite{Mead1990, Ribar2021, mendolia2026}. Reproducing excitability with an interpretable dynamical system has therefore become a recurring goal in mathematical neuroscience~\cite{Izhikevich2006}, nonlinear circuit design~\cite{Ribar2019}, and, more recently, control~\cite{Fyon2023, Cathcart2024, Bainier2025}.

Neuronal excitability is not a single behavior but a taxonomy. Three excitability types are distinguished according to how repetitive firing appears as a constant stimulus is increased~\cite{Hodgkin1948, Franci2013}: type~I fires at an arbitrarily low rate just above threshold, type~II starts firing abruptly at a nonzero rate, and type~III fires once at stimulus onset and never repetitively. A parallel taxonomy sorts bursters, which alternate silent phases with trains of spikes, by the pair of bifurcations that start and stop the burst~\cite{Franci2014, Izhikevich2000, samengo2013linking}: square-wave bursting rests on a fold and a homoclinic, parabolic bursting on two saddle-node on invariant circle transitions, and elliptic bursting on a subcritical Hopf and a fold of limit cycles. In every case above, the underlying bifurcation structure explains the phenomenology. Reduced models exposing that structure transparently have therefore been useful~\cite{Izhikevich2006, Fyon2024}, and they enabled a dynamical system reading of excitability.

This bifurcation-based reading has also been interpreted through the lens of control theory via the mixed-feedback principle~\cite{Sepulchre2018,Sepulchre2019}. Mixed feedback is an excitability motif which consists of a fast positive feedback loop, which provides regenerative amplification, interconnected with a slower negative feedback loop, which terminates the spike and resets the system. That viewpoint has proved fruitful for analysis, for neuromorphic circuit design~\cite{mendolia2026}, and for control synthesis~\cite{Ribar2021}. At the neurophysiological level, it is well-grounded, since all excitable neurons are circuit realizations of the mixed-feedback motif. However, the mixed-feedback motif alone does not suffice to characterize excitability. It always assumes a timescale hierarchy between the positive and negative feedback, and requires a particular arrangement of the nonlinear activation functions. Like biological motifs generally, it is necessary but not sufficient to determine a unique, well-defined function~\cite{Alon2006, Payne2015}. Synthesizing excitable systems by design therefore calls for narrower, functionally sufficient motifs. Although it may capture a constrained repertoire of dynamics, a minimal motif is both easier to analyze and to implement.

A simple sufficient excitability motif has recently been proposed~\cite{Bainier2026_poster, bainier2block}. A single high-gain saturating nonlinearity is closed in positive feedback around a linear frequency selector that blocks the constant component of its input and transmits its variations. A static nonlinearity and positive feedback on a selected band of high frequencies suffice to obtain an excitable primitive. The selector alone carries both roles that the mixed-feedback motif assigns to two separate paths. This is what makes the reduction possible. The selector is also a transfer function, so the system stays open to standard frequency-domain tools, and its behavior can be read from a pole placement and a Nyquist diagram.

This paper asks \emph{which bursting classes such a frequency-based primitive can produce, and what decides between them?} Burst classes rest on different bifurcations, so they respond differently to inputs~\cite{Izhikevich2000}, and neuromodulation switches neurons between them to change what they compute~\cite{Fyon2026a}. A minimal primitive is therefore only useful as a design element if we know which classes it can reach. We analyze the primitive with two elementary selectors, a first-order high-pass and a second-order resonant band-pass, and derive for each, with frequency-domain tools, the onset of firing, rest and firing bistability, and the bursting classes it supports. The nonlinearity decides whether the system fires and where the threshold sits. The selector decides whether rest and firing coexist. The contributions are as follows.
\begin{itemize}
    \item We show that, for any selector blocking constant inputs, the resting point is unique and set by the input alone, the feedback has zero mean at rest and during firing, and the nonlinearity alone sets the firing threshold.
    \item We show that a unique resting point rules out square-wave bursting. Without bistability, as with the first-order high-pass filter, no linear slow feedback with a single state can produce bursting either. A second slow oscillator is then needed, and it gives a parabolic burster.
    \item We show that a resonant selector gives type~II excitability, with an onset frequency set by the selector alone, and we derive the gain above which the onset becomes subcritical. Above that gain, rest and firing coexist, and a single slow negative feedback loop exploits this bistability to produce elliptic bursting.
\end{itemize}

The paper is organized as follows. Section~\ref{sec:model} defines the two-block motif. Section~\ref{sec:generic} establishes the properties of the motif that are shared across all frequency selectors. Section~\ref{sec:H0} treats the first-order high-pass selector, its monostability, and parabolic bursting. Section~\ref{sec:H1} treats the second-order resonant band-pass filter, its bistability window, and elliptic bursting. Section~\ref{sec:discussion} collects the comparison and its caveats, and Section~\ref{sec:conclusion} concludes.

\emph{Notation.} Signals are functions of time $t$, and $s$ is the Laplace variable; we write $w = H(s)\,y$ for the action of the filter $H$ on the signal $y$. For a rational transfer function, $H(\infty)$ is the limit of $H(s)$ as $|s| \to \infty$. Primes denote derivatives, $\langle \cdot \rangle$ the average of a periodic signal over one period, $\triangleq$ a definition, and a star a quantity evaluated at the equilibrium.  Fast quantities carry the subscript $\text{f}$ and slow ones the subscript $\text{s}$, with time constants $\tau_\text{f} \ll \tau \ll \tau_\text{s}$.

\section{The two-block setup}\label{sec:model}

\begin{figure}
    \vspace{0.2cm}
    \centering
    \includegraphics[width=\linewidth]{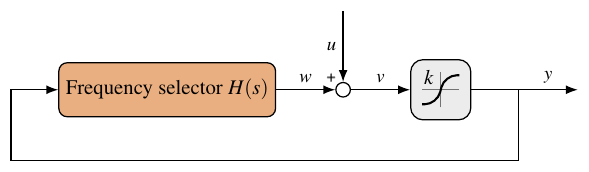}
    \caption{The two-block primitive. A monotone nonlinearity of slope $k$ at the origin is closed in positive feedback around a linear frequency selector $H(s)$ with $H(0)=0$. Adapted from~\cite{Bainier2026_poster}.}
    \label{fig:blockdiagram}
\end{figure}

The two-block setup is the positive feedback interconnection of a static nonlinearity and a linear frequency selector, as depicted in Fig.~\ref{fig:blockdiagram}:
\begin{equation} \label{eq:loop}
    y = \sig_k(v), \qquad v \triangleq u + w, \qquad w = H(s)\, y.
\end{equation}
Here, $u$ is the external input, $v$ and $y$ are the input and the output of the nonlinearity, $w$ is the feedback signal, and $H(s)$ the frequency selector.

\subsection{The nonlinearity}

The nonlinearity $\sig_k$ is monotone increasing, bounded, and odd, with a slope maximal at the origin, so that
\begin{equation}
    \sig_k'(0) = k ,
\end{equation}
and nonincreasing away from it. Throughout, \emph{nonlinearity} refers to this class. With the high-pass selector, we keep the saturation $\sat_k(v) = \tfrac{1}{2}\left(\left|kv+1\right| - \left|kv-1\right|\right)$ from the original study~\cite{Bainier2026_poster}, because it gives closed-form firing rates. With the band-pass selector, we use $\tanh(kv)$, because the criticality of the onset depends on the third derivative of the nonlinearity (Section~\ref{sec:H1}), which vanishes for the saturation. A smooth nonlinearity also gives type~II excitability with the high-pass selector, but rest and firing still never coexist~\cite{Bainier2026_poster}, which is the property that matters for bursting.

\subsection{The frequency selector}

\begin{assumptionx} \label{as:dc}
The selector blocks the constant component of its input, $H(0) = 0$.
\end{assumptionx}

Without Assumption~\ref{as:dc}, the constant component would be fed back and the loop would amplify rather than excite. Under Assumption~\ref{as:dc}, the feedback contributes nothing at steady state, so the primitive has a well-defined resting state set by the input alone. This paper investigates two cases.

\subsubsection{First-order high-pass, $H_\text{hp}$}
The high-pass selector has order one and relative degree zero, hence a direct feedthrough,
\begin{equation} \label{eq:H0}
    H_\text{hp}(s) = \frac{\tau s}{\tau s + 1}, \quad \text{with } H_\text{hp}(\infty) = 1,
\end{equation}
with the state-space form
\begin{equation} \label{eq:H0ss}
    \tau \dot{x} = -x + y, \qquad x \triangleq y - w .
\end{equation}
This is the original selector of~\cite{bainier2block}. Its unit feedthrough makes the output equation implicit, $y = \sig_k(u + y - x)$, hence multi-valued for $k>1$. The loop is made causal with a low-pass of time constant $\tau_\text{f} \ll \tau$ that is neglected ($\tau_\text{f} \to 0$) in the analysis, as any physical realization naturally does.

\subsubsection{Second-order resonant band-pass, $H_\text{r}$}
The resonant selector has order two and relative degree one, hence no
feedthrough,
\begin{equation} \label{eq:H1}
    H_\text{r}(s) = \frac{(\omega_n/Q)\,s}{s^2 + (\omega_n/Q)\,s + \omega_n^2}, \quad \text{with } H_\text{r}(\infty) = 0,
\end{equation}
with $\omega_n = 2\pi f_{\text{c}}$ the center frequency and $Q$ the quality factor, and the state-space form
\begin{subequations} \label{eq:H1ss}
\begin{align}
    \dot x_1 &= x_2, \\
    \dot x_2 &= -\omega_n^2 x_1 - \tfrac{\omega_n}{Q} x_2 + y, \\
    w &= \tfrac{\omega_n}{Q} x_2 .
\end{align}
\end{subequations}
Here, the output equation is explicit, so no regularization is required. We assume throughout that $Q > 1/2$, so that the selector is resonant.

\section{What holds for both selectors}\label{sec:generic}

\subsection{The resting point is unique}

\begin{lemma}[Resting point] \label{lem:rest}
Under Assumption~\ref{as:dc}, system~\eqref{eq:loop} has, for every constant input $u$, the unique equilibrium
\begin{equation}
    w^{*} = 0, \qquad y^{*} = \sig_k(u),
\end{equation}
so the operating point of the nonlinearity is the input itself, $v^{*} = u$.  Linearizing the loop at that equilibrium gives the characteristic equation $1 - \sig_k'(u) H(s) = 0$, where the loop gain
\begin{equation} \label{eq:loopgain}
    \sig_k'(u) \in [0, k]
\end{equation}
has maximum $k$, attained at $u = 0$; for $\sig_k(v) = \tanh(kv)$ it evaluates to $k\sech^2(ku)$.
\end{lemma}

\begin{proof}
At an equilibrium, the signals are constant, so $w = H(0) y = 0$ by Assumption~\ref{as:dc} and~\eqref{eq:loop} reduces to $y = \sig_k(u)$, which fixes $y$ uniquely. Perturbing~\eqref{eq:loop} around that point and eliminating $w$ gives the characteristic
equation $1 - \sig_k'(v^{*}) H(s) = 0$, and $v^{*}=u$.
\end{proof}

As the input sweeps $\mathbb{R}$, the operating point sweeps the whole nonlinearity, so the system can exhibit exactly the slopes of the nonlinearity as loop gains at rest. The selector never shifts that operating point. The nonlinearity therefore fixes which loop gains are reachable.

\begin{lemma}[Zero-mean feedback] \label{lem:blind}
Under Assumption~\ref{as:dc}, the feedback signal $w$ has zero mean at rest and on any periodic orbit of~\eqref{eq:loop}.
\end{lemma}

\begin{proof}
On a periodic orbit, the filter is in periodic steady state --- both selectors are stable --- so $\langle w \rangle = H(0)\,\langle y \rangle = 0$ by Assumption~\ref{as:dc}, whatever the mean of $y$. At rest, $w^{*}=0$ by Lemma~\ref{lem:rest}.
\end{proof}

Lemma~\ref{lem:blind} constrains the slow feedback loops of Sections~\ref{sec:H0burst} and~\ref{sec:elliptic}. A slow branch that senses $w$ is \emph{blind}: it reads the same zero mean whether the system rests or fires.

\subsection{The rheobase belongs to the nonlinearity}

Both selectors lose local stability at the same place.

\begin{proposition}[Local stability] \label{prop:split}
For the primitive~\eqref{eq:loop} with either selector, the threshold at which rest loses local stability depends only on the nonlinearity and is given by $\sig_k'(u)=1$.
\end{proposition}

\begin{proof}
By Lemma~\ref{lem:rest}, the linearization reads $1 - \sig_k'(u) H(s) = 0$ at the operating point $v^{*}=u$. For $H_\text{hp}$, it has the single real root
\begin{equation} \label{eq:H0root}
    s = -\frac{1}{\tau\,\left(1-\sig_k'(u)\right)} ,
\end{equation}
which is negative exactly for $\sig_k'(u)<1$. For $H_\text{r}$, the characteristic polynomial is
\begin{equation} \label{eq:H1poly}
    s^2 + \frac{\omega_n}{Q}\,\left(1-\sig_k'(u)\right)\,s + \omega_n^2 ,
\end{equation}
whose coefficients are all positive, and which is therefore Hurwitz exactly for $\sig_k'(u)<1$. Overall, both boundaries coincide with $\sig_k'(u)=1$.
\end{proof}

Solving $\sig_k'(u)=1$ for $\tanh$ gives the rheobase
\begin{equation} \label{eq:rheobase}
    |u_\text{th}| = \frac{1}{k}\,\artanh\left( \sqrt{1-\frac{1}{k}} \right) ,
\end{equation}
which exists only for $k>1$. For the saturation $\sat_k(v)$, it is $1/k$. Expression~\eqref{eq:rheobase} is not monotone in $k$: it rises to a maximum at $k_{\max}\approx 1.724$ and then decays as $\ln(4k)/2k$.

The remainder of the paper characterizes how the selector decides whether rest and firing can coexist, and hence which bursting classes are available.

\subsection{A frequency-domain stability analysis}

\begin{figure}
    \vspace{0.2cm}
    \centering
    \scalebox{0.95}{%=========================================================== (a) sigmoid
\begin{tikzpicture}[scale=1.6,>={Stealth[length=4pt]},
                    baseline=(current bounding box.center)]

  % sector [0,k] : the region between y=0 and y=kv
  \fill[gray!10] (0,0) -- (0.525,1.05) -- (1.0,1.05) -- (1.0,0) -- cycle;
  \fill[gray!10] (0,0) -- (-0.525,-1.05) -- (-0.95,-1.05) -- (-0.95,0) -- cycle;

  % axes
  \draw[->,gray!55] (-1.02,0) -- (1.12,0) node[below=1pt,black]{$v$};
  \draw[->,gray!55] (0,-1.14) -- (0,1.20) node[above=0pt,black]{$\varphi_k(v)$};

  % upper edge of the sector, y = kv, which is also the tangent at v=0
  \draw[gray!65,line width=0.5pt] (-0.525,-1.05) -- (0.525,1.05);
  \node at (0.70,0.15) {sector $[0,k]$};

  % the sigmoid, tanh(kv) with k=2
  \draw[very thick,domain=-0.95:1.0,samples=120,smooth]
        plot (\x,{(exp(4*\x)-1)/(exp(4*\x)+1)});

  % drop lines to the input axis: the operating point is v*=u
  \draw[gray!45,dotted] (0.441,0) -- (0.441,0.707);
  \draw[gray!45,dotted] (0.658,0) -- (0.658,0.866);
  \node[below=2pt] at (0.441,0) {$u_H$};

  % markers
  \mkcirc{0}{0}
  \mksq{0.441}{0.707}
  \mktri{0.658}{0.866}

  % legend, in the empty lower right
  \mkcirc{0.34}{-0.42} \node[right=3pt] at (0.40,-0.42) {$\varphi_k'(u)=k$};
  \mksq{0.34}{-0.73}   \node[right=3pt] at (0.40,-0.73) {$\varphi_k'(u)=1$};
  \mktri{0.34}{-1.04}  \node[right=3pt] at (0.40,-1.04) {$\varphi_k'(u)<1$};

\end{tikzpicture}
\hspace{1mm}
%=========================================================== (b) Nyquist
\begin{tikzpicture}[scale=1.6,>={Stealth[length=4pt]},
                    baseline=(current bounding box.center)]

  % critical points 1/a, one vertical per operating point
  \draw[gray!55,dashed,line width=0.4pt] (0.6,-0.68) -- (0.6,0.70);
  \draw[gray!55,dashed,line width=0.4pt] (1.0,-0.68) -- (1.0,0.70);
  \draw[gray!55,dashed,line width=0.4pt] (1.5,-0.68) -- (1.5,0.70);

  % axes
  \draw[->,gray!55] (-0.26,0) -- (1.92,0) node[below=1pt,black]{$\mathrm{Re}$};
  \draw[->,gray!55] (0,-0.78) -- (0,0.84) node[above=0pt,black]{$\mathrm{Im}$};

  % circle of diameter [0,1] : upper half, common to both selectors
  \draw[very thick,
        decoration={markings,mark=at position 0.55 with
                    {\arrow{Stealth[length=7pt,width=6pt]}}},
        postaction={decorate}]
        (0,0) arc[start angle=180,end angle=0,radius=0.5];

  % lower half, traversed by H1 only
  \draw[very thick,dashed,
        decoration={markings,mark=at position 0.55 with
                    {\arrow{Stealth[length=7pt,width=6pt]}}},
        postaction={decorate}]
        (1,0) arc[start angle=0,end angle=-180,radius=0.5];

  \node[above=1pt] at (0.5, 0.50) {$H_\mathrm{hp},H_\mathrm{r}$};
  \node[below=1pt] at (0.5,-0.50) {$H_\mathrm{r}$ only};
  \node[anchor=north east,inner sep=2pt] at (0,0) {$\omega=0$};

  % the three critical points
  \mkcirc{0.6}{0}
  \mksq{1.0}{0}
  \mktri{1.5}{0}
  \node[below=4pt] at (0.6,0) {$1/k$};
  \node[below=4pt] at (1.05,0) {$1$};
  \node[below=4pt] at (1.5,0) {$\frac{1}{\varphi_k'(u)}$};

  % direction of travel of the critical point
  \draw[->,gray!70,line width=0.5pt] (0.5,-1.1) -- (1.5,-1.1);
  \node[below=0pt] at (1.0,-1.1) {$|u|$ increasing};

\end{tikzpicture}}
        \caption{Sector and Nyquist loci. \textbf{Left}: the nonlinearity and the sector $[0,k]$ that contains it. Sweeping $u$ sweeps the slope over the whole range $[0,k]$. \textbf{Right}: the Nyquist loci of both selectors lie on the circle of diameter $[0,1]$. Each vertical at $\Real\{H\}=1/\sig_k'(u)$ is the corresponding circle-criterion boundary, and the locus lies strictly to the left of all of them.}
    \label{fig:nyquist}
\end{figure}
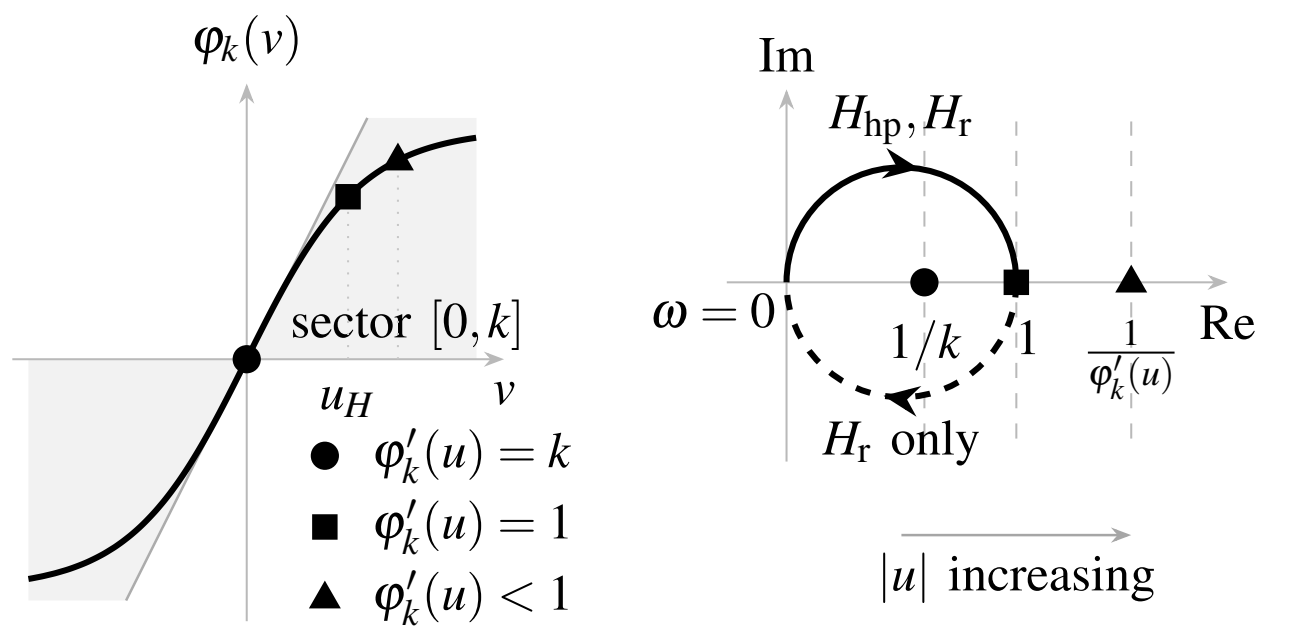

Since the primitive is a Lur'e system, the global stability of the system can be read directly off a Nyquist plot using the circle criterion, and the same picture carries the threshold and the frequency at which the system can oscillate. Both selectors satisfy the following identity, and $H_\text{r}$ does so for every $Q$:
\begin{equation} \label{eq:circle}
    |H(j\omega)|^2 = \Real\{H(j\omega)\} \quad \Longleftrightarrow \quad \left| H(j\omega) - \tfrac{1}{2} \right| = \tfrac{1}{2},
\end{equation}
so both Nyquist loci lie on the circle of diameter $[0,1]$ and differ only in how they travel along it (Fig.~\ref{fig:nyquist}): $H_\text{hp}(j\omega)$ climbs the upper half and approaches the point $1$ only as $\omega \to \infty$, whereas $H_\text{r}(j\omega)$ reaches it at $\omega=\omega_n$ and returns to the origin along the lower half. That point is where the system sustains an oscillation at unit gain; one selector reaches it at a finite frequency, and the other only at infinite frequency. The same picture also bounds the loop gain below which nothing at all can happen.

\begin{proposition}[Global stability below unit gain] \label{prop:global}
If $\sig_k'(0) < 1$, the resting point of~\eqref{eq:loop} is globally asymptotically stable for every constant input and for either selector. No limit cycle or other invariant set exists besides the resting point.
\end{proposition}

\begin{proof}
By Lemma~\ref{lem:rest}, the operating point sweeps the whole nonlinearity as the input sweeps the real axis, so the nonlinearity seen from any resting point lies in the sector $[0,k]$ with $k=\sig_k'(0)$. The circle criterion~\cite{Khalil2002} for the positive feedback interconnection requires $\Real\{H(j\omega)\}  < 1/k$ for all $\omega$, and, by~\eqref{eq:circle}, the supremum of $\Real\{H(j\omega)\} $ is $1$. The condition therefore holds for every frequency if and only if $k<1$.
\end{proof}

Below the bound $k=1$, the system is a stable filter that never fires repetitively. Above the bound, the system is in tonic spiking at $u=0$, since $\sig_k'(0)=1$ is also the bifurcation of Proposition~\ref{prop:split}. \emph{We assume $k>1$ throughout the remainder of the paper.}

\section{Parabolic bursting with the high-pass selector}
\label{sec:H0}

\subsection{The firing band}

The results of this subsection are established in~\cite{Bainier2026_poster} and recalled here for later use.
With $H_\text{hp}$, the closed-loop pole~\eqref{eq:H0root} is real and goes to infinity at $\sig_k'(u)=1$. With the saturation, stability is therefore lost through a fold of the algebraic output map~\eqref{eq:loop} rather than at a finite frequency, and the firing rate starts from zero, which is type~I excitability. Two consequences are used below. First, the rate is available in closed form,
\begin{equation} \label{eq:frequency}
    t_\pm(u) = \tau \ln \frac{2k-1\mp k|u|}{1\mp k|u|}, \qquad
    f(u) = \frac{1}{t_+(u) + t_-(u)},
\end{equation}
maximal at zero input, where $f_{\max} = \left(2\tau\ln(2k-1)\right)^{-1}$. Second, since $\sig_k'(u)=\sig_k'(-u)$, the condition $\sig_k'(u)>1$ defines a symmetric band $|u|<u_\text{th}$ at both edges of which $f$ vanishes: the system fires within a window and falls silent beyond either edge, with the same threshold at each edge.

\subsection{Bursting with the high-pass selector}
\label{sec:H0burst}

Any slow feedback branch added to~\eqref{eq:loop} is subject to two structural constraints, whatever the selector. First, by Lemma~\ref{lem:blind}, a linear slow branch cannot take $w$ as input: the trace of $w$ does distinguish rest from firing, but a slow branch responds to the mean, which vanishes in both regimes. Therefore, it must sense the output $y$. Second, the linear slow branch output must enter the loop at the input port of the nonlinearity rather than at the input port of the selector. Indeed, an input $u_{\text{c}}$ injected at the selector node gives $w = H(0)(y + u_{\text{c}}) = 0$ at any equilibrium, so the operating point remains $v^{*} = u$ and the loop gain~\eqref{eq:loopgain} is independent of $u_{\text{c}}$. An input at the selector port drives transients but cannot modulate excitability: no constant $u_{\text{c}}$ ever brings the system to threshold.

\begin{proposition}[No square-wave bursting] \label{prop:square}
Square-wave bursting, which requires a fast subsystem that is bistable over a range of the slow variable, cannot be produced by adding linear slow dynamics to~\eqref{eq:loop}.
\end{proposition}

\begin{proof}
By the second constraint above, a slow variable can only enter at the port of the nonlinearity, where it shifts the effective input. On the fast timescale, it is constant, so the fast subsystem is~\eqref{eq:loop} driven by a constant input, which has a unique equilibrium by Lemma~\ref{lem:rest}. The output map is multivalued for $k>1$, but that multivaluedness is what produces the spike; the fast subsystem including the filter state has a unique equilibrium at every value of the slow variable, and is therefore never bistable.
\end{proof}

Proposition~\ref{prop:square} leaves exactly two openings for bursting: drive the system across its own firing threshold with a slow signal, treated below, or exploit a coexistence between an equilibrium and a limit cycle (Section~\ref{sec:elliptic}), which requires a subcritical onset. Restoring the bistability of the fast subsystem itself would take a nonmonotone nonlinearity. Here, only the first opening is available, since the system has no memory of whether it was firing. A burster cannot be built by making the system remember; it has to be built by moving the system across the firing band $|u| < u_\text{th}$.

\subsubsection{Exogenous drive gives parabolic bursting}

Let $u(t)$ be periodic and slow, of period $T_\text{slow} \gg \tau$. Bursts are the passages through the band, and since $f$ vanishes at both edges, the spike rate slows at burst onset and offset, giving a parabolic structure~\cite{ErmentroutKopell1986}. The spike count per burst follows the quasistatic prediction
\begin{equation} \label{eq:spikecount}
    N = \int_\text{burst} f\big(u(t)\big)\,dt ,
\end{equation}
with $f$ given by~\eqref{eq:frequency}. A symmetric drive gives two bursts per slow period. One burst per period requires entering and leaving the band from the same side, that is $-u_\text{th} < u_0 - A < u_\text{th} < u_0 + A$ for a sinusoid of offset $u_0$ and amplitude $A$.

\subsubsection{No single linear slow branch can close the burst}

The next step is to replace the external drive by an internal slow branch that moves the operating point of the system. A linear branch fails, whether it senses $w$ or $y$.

\begin{proposition}[No bursting from a single linear slow branch] \label{prop:noslow}
Under Assumption~\ref{as:dc} and with the selector $H_\text{hp}$, no additional linear slow branch carrying a single state produces bursting in the slow limit.
\end{proposition}

\begin{proof}
By Proposition~\ref{prop:square}, the fast subsystem has a unique equilibrium at every frozen value of the slow variable. With $H_\text{hp}$, rest and repetitive firing never coexist either (Section~\ref{sec:H0mono}). What a slow branch averages is therefore a single-valued function of its own state, and the added branch reduces, in the slow limit, to a scalar differential equation without hysteresis. Such a flow converges monotonically to a fixed point, whereas bursting requires the slow variable to cycle. Only the location of that fixed point changes from one branch to the next.

Sensing $w$ fails for a second and independent reason, given at the beginning of this subsection. The argument of Lemma~\ref{lem:blind} needs only that a stable filter passes the mean of its input scaled by its DC gain, so the same blindness affects any branch $G$ placed downstream of the selector or in parallel with it, and any branch with $G(0)=0$ placed anywhere. Sensing $y$ escapes Lemma~\ref{lem:blind}, since the output carries a nonzero mean, but the scalar argument still applies. In the slow limit, a linear branch of DC gain $G(0)$ returns $G(0)\,Y(u)$, where $Y(u) \triangleq \langle y \rangle$ is the mean output at a frozen input; sensing the filter state $x$ is the same case, since the low-pass has unit DC gain and $\langle x \rangle = Y(u)$. For the saturation $\sat_k(v)$, $Y$ is continuous and increasing, and known in closed form: $(t_+(u)-t_-(u))/(t_+(u)+t_-(u))$ inside the band, $\sign(u)$ outside. Closing the branch in negative feedback gives the scalar flow $\tau_\text{s}\dot z = -z + Y(u-gz)$, $g>0$, whose right-hand side has derivative $-1-g\,Y'(u-gz) \le -1$. The fixed point is unique and globally attracting, and the system either rests or fires tonically.
\end{proof}

\begin{remark}
A nonlinear branch can tell rest from firing, yet its averaged flow is still scalar. Rectifying the feedback, for instance, gives $\langle |w| \rangle = 0$ at rest and $\langle |w| \rangle > 0$ during firing, but the averaged flow stays single-valued. Escaping the scalar argument would require a branch carrying hysteresis of its own~\cite{Franci2018}, hence at least a second nonlinearity and one more block.
\end{remark}

\begin{figure}[!t]
    \vspace{0.2cm}
    \centering
    \includegraphics[width=\linewidth]{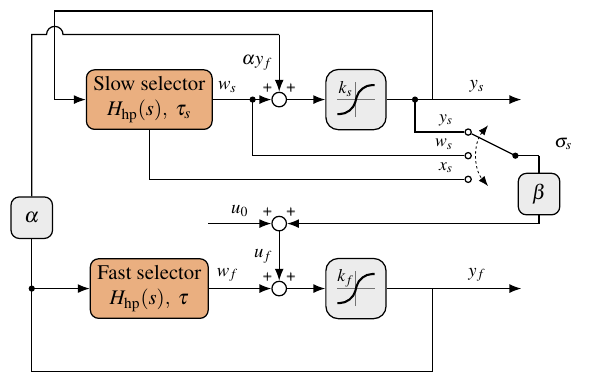}
    \caption{The cascade~\eqref{eq:cascade}. Two copies of the primitive of Fig.~\ref{fig:blockdiagram} run at $\tau_\text{s} \gg \tau$, the slow one biased inside its own firing band. The switch selects which signal $\sigma_\text{s}$ of the slow loop drives the fast one, and the gain $\beta$ and the bias $u_0$ place the resulting sweep with respect to the firing band of the fast loop. The return gain $\alpha$ modulates the burst: raising it reduces the burstiness.}
    \label{fig:cascadediagram}
\end{figure}

\subsubsection{A second copy of the primitive is a minimal burster}

What does work is to replace the external slow drive by a second copy of the same two blocks, running at $\tau_\text{s} \gg \tau$ and biased inside its own firing band so that it fires slowly~\cite{Ribar2017}, as in Fig.~\ref{fig:cascadediagram}. Its own relaxation cycle then supplies the sweep:
\begin{subequations} \label{eq:cascade}
\begin{align}
    \tau_\text{s}\,\dot x_\text{s} &= -x_\text{s} + y_\text{s}, & y_\text{s} &= \sig_{k_\text{s}}(\alpha y_\text{f} + y_\text{s} - x_\text{s}), \\    \tau\,\dot x_\text{f} &= -x_\text{f} + y_\text{f}, & y_\text{f} &= \sig_{k_\text{f}}(u_0 + \beta \sigma_\text{s} + y_\text{f} - x_\text{f}),
\end{align}
\end{subequations}
where $\sigma_\text{s} \in \{x_\text{s}, w_\text{s}, y_\text{s}\}$ is the selected signal of the slow loop to drive the fast loop, and the gain $\alpha$ returns the fast output to the slow loop (Fig.~\ref{fig:cascadediagram}). The slow drive is now produced by the model itself rather than imposed from outside, with no new block type and two dynamic states in total, against the one fast and two slow states of a textbook parabolic burster~\cite{Izhikevich2006}. The burst boundaries are still set by the slow loop, since the spikes of the fast loop do not terminate the burst on their own. Raising the gain $\alpha$ reduces the burstiness, but the switch position decides how: it lowers the interburst frequency for $\sigma_\text{s} = x_\text{s}$, and the spike count per burst for the other two.

\begin{figure}
    \vspace{0.2cm}
    \centering
    \includegraphics[width=\linewidth]{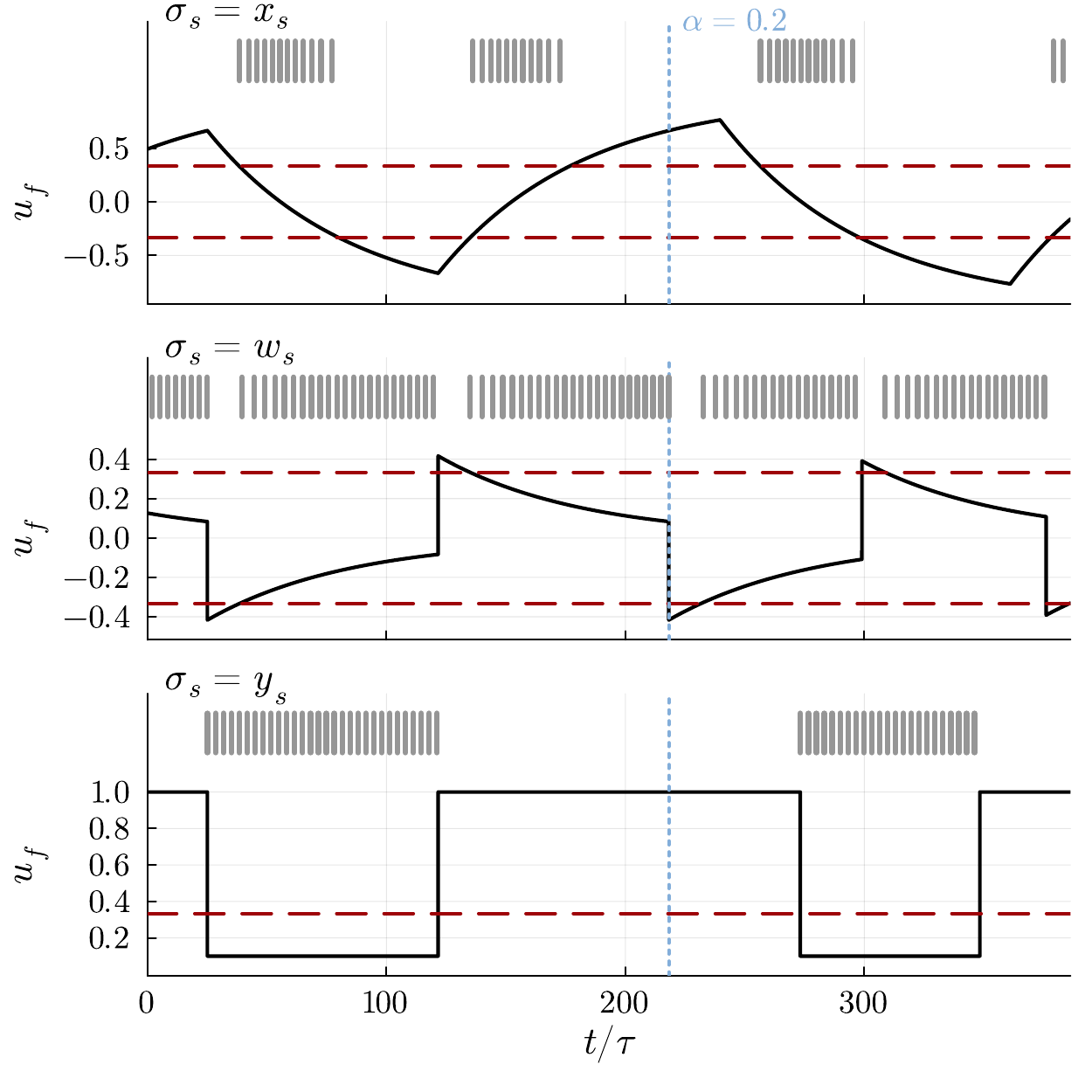}
    \caption{The cascade~\eqref{eq:cascade} at $k_\text{s}=k_\text{f}=3$, $\tau_\text{s}=60$ and $\tau=1$. Top to bottom, the switch selects $\sigma_\text{s} = x_\text{s}$ with $\beta=1$, $u_0=0$, then $w_\text{s}$ with $\beta=0.25$, $u_0=0$, then $y_\text{s}$ with $\beta=0.45$, $u_0=0.55$. Black is the effective input $u_\text{f} = u_0+\beta\sigma_\text{s}$ of the fast loop, red dashed are the edges $\pm u_\text{th}$ of the firing band, gray ticks are the spike times. In every case, the burst is the passage of the black curve between the dashed lines. At the blue dashed line, $\alpha$ steps from $0$ to $0.2$.}
    \label{fig:cascade}
\end{figure}

The position of the switch in Fig.~\ref{fig:cascadediagram} selects the burst class, as shown in Fig.~\ref{fig:cascade}. Each of the three crosses the firing band with a different shape. Note that $w_\text{s}$ is usable here, unlike in the previous subsection, because it is routed directly to the fast loop instead of being averaged by a slow branch. Selecting the filter state $x_\text{s}$ gives a smooth ramp across the band and therefore a parabolic burst. Selecting the feedback signal $w_\text{s}$ gives a decaying plateau and an accelerating burst. Selecting the output $y_\text{s}$ puts the operating point straight into the middle of the band and straight back out, so burst onset and offset are abrupt and the intra-burst rate is flat. It looks like square-wave bursting even though the mechanism underneath is a threshold crossing on each side. This position requires the bias $u_0$, since the input alternates between $u_0+\beta$ and $u_0-\beta$ and bursting needs exactly one of the two inside the band, that is $|u_0-\beta| < u_\text{th} < |u_0+\beta|$. Here, the visual signature and the bifurcation mechanism are decoupled, so bursters cannot be classified from traces alone.

\subsection{The high-pass selector is monostable}
\label{sec:H0mono}

With $H_\text{hp}$, the characteristic equation $\tau\,\left(1-\sig_k'(u)\right)\,s + 1 = 0$ carries a single real root, but this only describes rest locally. Since $\sig_k'$ is maximal at the origin and nonincreasing away from it, the output map $y=\sig_k(u+y-x)$ has exactly two stable branches, and a limit cycle must visit both. If rest is stable, the equilibrium lies on one of them and the slow flow of $x$ converges to it, so no limit cycle survives. The onset depends on the nonlinearity: the rate starts from zero with the saturation, and from a finite value with a smooth sigmoid, which is type~II excitability~\cite{Bainier2026_poster}. In both cases, rest and repetitive firing never coexist, and the system is monostable at every input. The second opening left by Proposition~\ref{prop:square} is therefore closed, and elliptic bursting is out of reach with $H_\text{hp}$, whatever the nonlinearity. With a resonant second-order band-pass, rest and firing do coexist above a computable gain, which is the subject of Section~\ref{sec:H1}.

\section{Elliptic bursting with the band-pass selector}
\label{sec:H1}

The high-pass selector never lets rest and firing coexist. We therefore take the resonant second-order form~\eqref{eq:H1} with $Q > 1/2$. Its phase runs from $+90^\circ$ to $-90^\circ$ and therefore crosses zero at $\omega = \omega_n$, where the gain is exactly one.

\subsection{Type~II excitability}

Linearizing~\eqref{eq:H1ss} around the equilibrium gives the characteristic polynomial~\eqref{eq:H1poly}. Three consequences follow.

\emph{Threshold.} By Proposition~\ref{prop:split}, the rheobase is again~\eqref{eq:rheobase}, identical to the threshold with $H_\text{hp}$.

\emph{Onset frequency.} At $\sig_k'(u)=1$, the polynomial~\eqref{eq:H1poly} reduces to $s^2 + \omega_n^2$, so
\begin{equation} \label{eq:onsetfreq}
    \omega_\text{Hopf} = \omega_n
\end{equation}
independently of $k$ and of $Q$. Spiking therefore starts at a nonzero rate, fixed by the selector alone, which is the defining signature of type~II excitability.

\emph{Subthreshold resonance.} Below threshold, the damping ratio is $\zeta = \left(1-\sig_k'(u)\right)/2Q$, which the standing assumption $Q>1/2$ keeps below one at every input.
The resting system therefore rings at a frequency $\omega_n\sqrt{1-\zeta^2}$ that varies with the input. This preferred frequency is the observable signature of resonance, and the single real pole of $H_\text{hp}$ has none.

\subsection{Soft and hard onset}
Whether the onset is soft or hard is decided by the nonlinearity. Shifting the origin to the equilibrium with $\xi = x_1 - \tanh(ku)/\omega_n^2$ and $\eta = x_2$, expanding the nonlinearity to third order and setting $\omega_n=1$, which is only a time rescaling, gives
\begin{equation} \label{eq:normalform}
    \dot\xi = \eta, \qquad \dot\eta = -\xi + \mu\eta + b\,\eta^2 + c\,\eta^3 ,
\end{equation}
with
\begin{equation}
    \mu = \frac{\sig_k'(u)-1}{Q}, \quad b = \frac{\sig_k''(u)}{2Q^2},
    \quad c = \frac{\sig_k'''(u)}{6Q^3} .
\end{equation}
Averaging with $\eta \approx -A\sin t$, using $\langle \eta^2\rangle = A^2/2$, $\langle \eta^3 \rangle = 0$ and $\langle \eta^4 \rangle = 3A^4/8$, yields the amplitude equation
\begin{equation} \label{eq:amplitude}
    \dot A = \frac{\mu}{2}A + \frac{3c}{8}A^3 ,
\end{equation}
which is the canonical amplitude equation of a Hopf bifurcation, so that the sign of the third derivative of the nonlinearity at threshold decides the criticality. For $\tanh$, using $1-\tanh^2(ku_\text{th}) = 1/k$ at threshold,
\begin{equation} \label{eq:criticality}
    \sig_k'''(u_\text{th}) = 2k\,(2k-3) .
\end{equation}

\begin{proposition}[Criticality] \label{prop:criticality}
The Hopf bifurcation of the primitive~\eqref{eq:loop} with selector $H_\text{r}$ and $\sig_k(v) = \tanh(k v)$ is supercritical for $1<k<3/2$, and subcritical for $k>3/2$.
\end{proposition}

\begin{proof}
The nonzero branch of~\eqref{eq:amplitude} is $A^2 = -4\mu/3c$. The sign of $c$ decides. For $c<0$, the branch exists if $\mu>0$ and is stable, so the onset is supercritical. For $c>0$, it exists if $\mu<0$ and is unstable, so the onset is subcritical. Since $Q>0$, $c$ has the sign of $\sig_k'''(u_\text{th})$ in~\eqref{eq:criticality}: negative for $1<k<3/2$, positive for $k>3/2$, zero at $k=3/2$. There, the cubic term vanishes and higher orders decide.
\end{proof}

\begin{figure}[!t]
    \vspace{0.2cm}
    \centering
    \includegraphics[width=\linewidth]{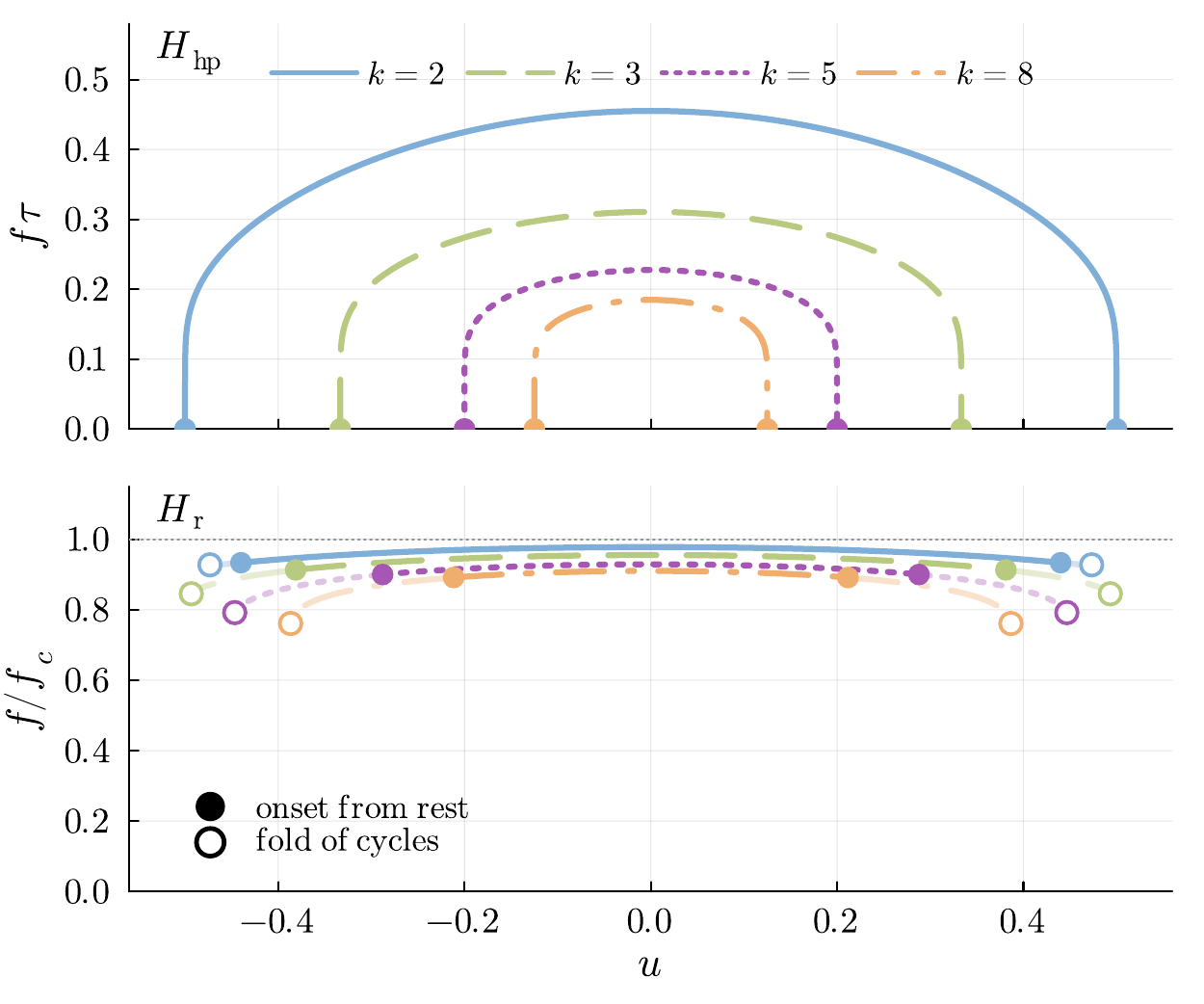}
    \caption{Rate-input curves at $Q=1$, for four gains $k$.
    \textbf{Top}: selector $H_\text{hp}$, with the saturation $\sat_k(v)$; thresholds differ from those in the bottom panel only because the two nonlinearities differ.
    \textbf{Bottom}: selector $H_\text{r}$ with $\tanh$. The dotted line is $f_{\text{c}}$. Filled markers give the onset, open markers the fold of cycles. Both are located with a decay test on the late amplitude and a regularity test on the measured periods. Both sit below $f_{\text{c}}$ because the subcritical cycle is relaxation-like. Both also drift downward as $k$ grows, by a decreasing amount.}
    \label{fig:fI}
\end{figure}

\begin{figure*}[!t]
  \vspace{0.2cm}
  \centering
  \includegraphics[width=0.9\linewidth]{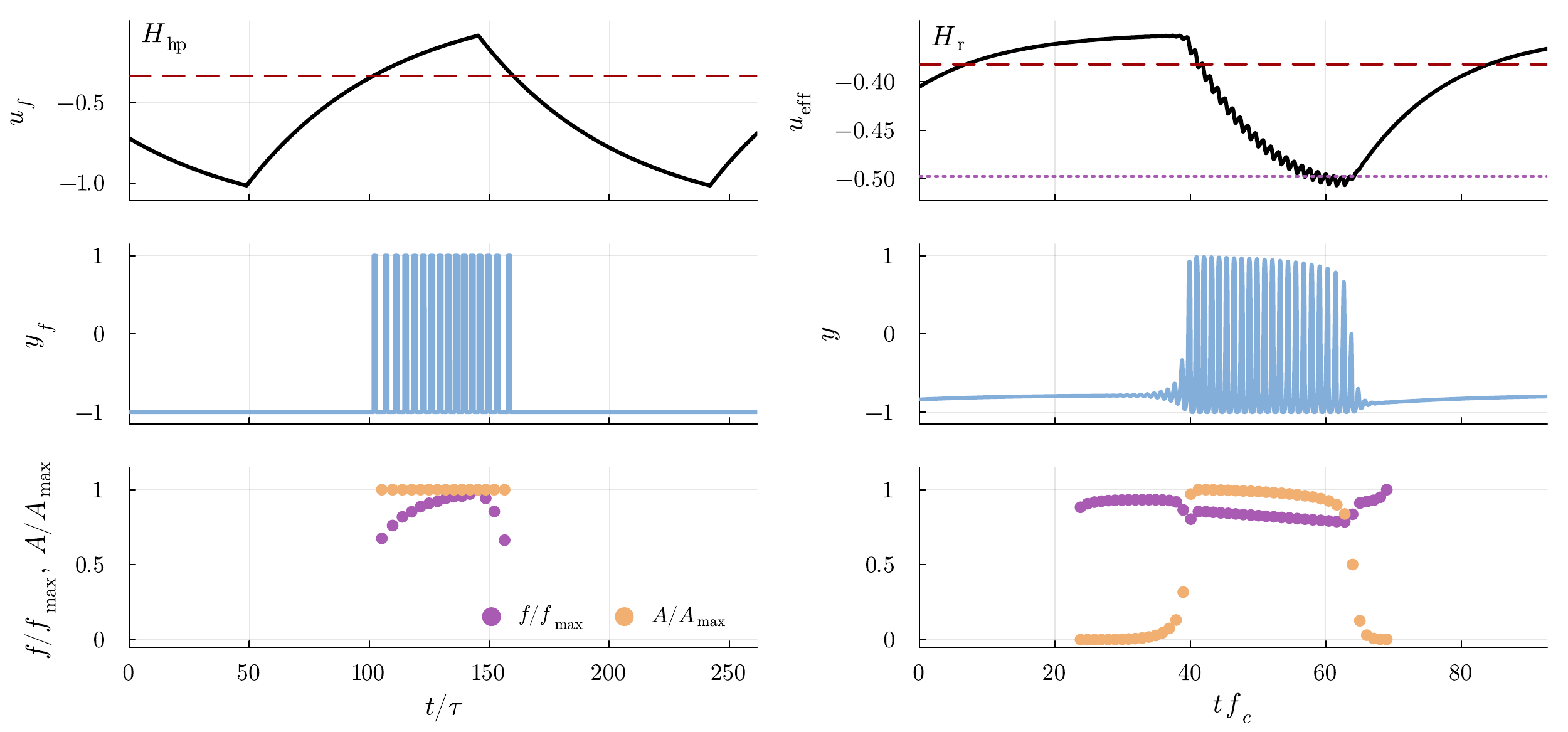}
  \caption{The selector chooses the burst class. \textbf{Left}: cascade~\eqref{eq:cascade} with the selector $H_\text{hp}$, switch position $\sigma_\text{s} = x_\text{s}$, at $k_\text{s}=k_\text{f}=3$, $\tau_\text{s}=60$, $\tau=1$, $\alpha=0$, $\beta=0.7$ and $u_0=-0.55$. \textbf{Right}: selector $H_\text{r}$ with the single slow state~\eqref{eq:slow}, at $k=3$, $Q=1$, $f_{\text{c}}=\SI{1}{\hertz}$, $\tau_\text{s}=20$, $u=-0.78$ and $g=0.55$, for which $|u_\text{eff}|$ swings over $[0.353, 0.507]$. Two slow periods are shown in each column, on the time axes $t/\tau$ and $t f_{\text{c}}$ respectively. \textbf{Top}: the effective input at the port of the nonlinearity, with the band edge $-u_\text{th}$ dashed on the left, and the subcritical Hopf dashed and the fold of limit cycles dotted on the right. \textbf{Middle}: the output of the system. \textbf{Bottom}: instantaneous rate (purple) and peak-to-peak amplitude (orange) measured over each interspike interval, each normalized to its own maximum over the window shown, so that the two columns are directly comparable.}
  \label{fig:comparison}
\end{figure*}

A supercritical onset produces spikes that emerge with infinitesimally small amplitude and grow as the input is pushed further, which is not an all-or-none event. A subcritical onset produces full-size spikes immediately, and leaves a range of inputs over which rest and spiking coexist. The latter is what a type~II neuron does, and it is why the initial condition matters when measuring a rate-input curve. A hard onset therefore requires $k>3/2$. Above that value, the spikes reach full amplitude at once, and the bistability window widens with $k$ before saturating. Fig.~\ref{fig:fI} compares the rate-input curves of the two selectors. With the saturation, the high-pass curve falls continuously to zero at the threshold. The type~II curves stay near $f_{\text{c}}$ and stop at a finite rate, and the stretch beyond the bifurcation exists only because the sweep is initialized off the equilibrium. That stretch is the bistability window, and its width is a direct measure of how hard the onset is.

\subsection{Elliptic bursting}
\label{sec:elliptic}

\begin{figure}[t!]
    \centering
    \includegraphics[width=\linewidth]{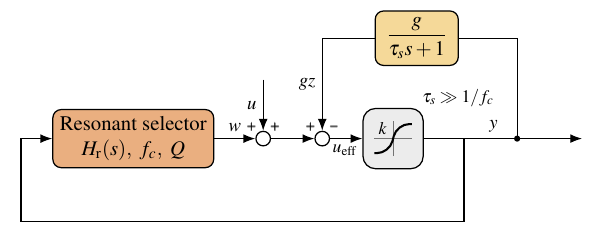}
    \caption{The elliptic burster~\eqref{eq:slow}. A single slow low-pass state, closed in negative feedback of gain $g$ around the resonant selector, sweeps the effective input $u_\text{eff}$ across the bistability window between the subcritical Hopf and the fold of limit cycles.}
    \label{fig:ellipticdiagram}
\end{figure}

For $k>3/2$, the bistability window of Proposition~\ref{prop:criticality} can be exploited by a slow feedback loop to achieve bursting. An averaged slow flow can be made hysteretic, which Proposition~\ref{prop:noslow} denied to the high-pass selector. By the two constraints of Section~\ref{sec:H0burst}, the slow branch must sense the output $y$ and must re-enter at the port of the nonlinearity. The plainest slow branch is therefore a first-order low-pass fed back negatively (Fig.~\ref{fig:ellipticdiagram}):
\begin{equation} \label{eq:slow}
    \tau_\text{s}\dot z = -z + y, \qquad u_\text{eff} = u - g z, \qquad \tau_\text{s} \gg 1/f_{\text{c}} .
\end{equation}
This construction adds no new nonlinearity.

The slow drive sweeps $u_\text{eff}$ back and forth across the bistability window, the spike amplitude within a burst grows and collapses, and the interburst phase is a subthreshold oscillation rather than pure silence. Bursting requires the sweep to reach both ends of that window and to stop there. This confines it to a thin diagonal strip in the $(u,g)$ plane, wedged between rest and tonic firing, whose width is the width of the bistability window.

The amplitude envelope distinguishes this from Section~\ref{sec:H0burst}. Here, the rate stays near $f_{\text{c}}$ and the amplitude varies within a burst. The contrast is shown in Fig.~\ref{fig:comparison}, and it comes from the two bifurcation mechanisms: threshold crossing modulates frequency, and Hopf modulates amplitude.

\addtolength{\textheight}{-0.5cm}

\section{Discussion}\label{sec:discussion}

Table~\ref{tab:summary} collects the results that distinguish the two frequency selectors. The nonlinearity fixes the rheobase, and together with the selector it shapes the onset: the high-pass gives type~I or type~II depending on the nonlinearity. What the selector decides is whether rest and firing can coexist at all, never with the high-pass and above a computable gain with the band-pass. This coexistence is what separates parabolic from elliptic bursting.

\begin{table}[t]
\caption{The selector decides whether rest and firing can coexist, and hence the burst class.}
\label{tab:summary}
\centering
\small
\setlength{\tabcolsep}{4pt}
\begin{tabular}{lll}
\toprule
                & $H_\text{hp}$               & $H_\text{r}$ \\
\midrule
order           & first               & second \\
relative degree & zero                & one \\
poles           & real                & complex pair ($Q > 1/2$) \\
nonlinearity used & saturation        & $\tanh$ \\
excitability    & I (II if $\tanh$) & II \\
bistability & never & for $k>3/2$ \\
burst class     & parabolic           & elliptic \\
modulated       & spike rate          & spike amplitude \\
\bottomrule
\end{tabular}
\end{table}

Two limitations are worth stating. Square-wave bursting is structurally unavailable, as a consequence of the uniqueness of the resting point (Proposition~\ref{prop:square}). And the elliptic regime occupies a narrow region of parameter space, whose width is inherited from the bistability window; mapping such regions is itself an active problem in neuroscience~\cite{Brandoit2026}.

Against a mixed-feedback realization, the primitive saves a block. Mixed feedback needs a second filter to separate the fast and the slow branch, whereas here a single selector carries both roles, and fewer blocks usually means fewer states to hold in a digital implementation. What that block count buys in an analog realization is still an open question.

The classification is a design rule as much as a taxonomy. A bank of $H_\text{r}$ filters tuned to different center frequencies detects narrow-band events directly in spikes, as event-based vibration monitoring or acoustic transient detection would require. The two selectors also encode information in different ways: the high-pass converts amplitude into a spiking rate, whereas the resonant band-pass pins the rate at $f_{\text{c}}$ and moves the information into timing and phase. Finally, the cascade of Section~\ref{sec:H0burst} is a burst generator with two slow-fast states, which is the building block of a central pattern generator~\cite{Fyon2026b}.

\section{Conclusions and future work}\label{sec:conclusion}

The nonlinearity of the primitive decides whether the primitive fires and where the threshold sits, and the frequency selector decides whether rest and firing can coexist. Between them, two elementary selectors already cover two of the three Hodgkin excitability types and two of the three standard bursting classes. The high-pass never makes rest and firing coexist, whatever the nonlinearity, so it only supports bursters that cross the firing band, such as the parabolic one. The resonant band-pass carries a complex pair of poles, and its subcritical Hopf opens the bistability window that an elliptic burster needs.

Some questions remain open. Which selectors allow rest and firing to coexist is uncharacterized beyond the two treated here. Moreover, moving a primitive between excitability types and burst classes is what neuromodulation does in biology~\cite{Fyon2026a}, and such retuning remains to be investigated. Finally, a subthreshold CMOS realization of both selectors would turn this classification into a design procedure.

\appendices
\section*{Appendix}

All code can be found online on GitHub: \url{https://github.com/arthur-fyon/ECC_2027_Fyon}.

\bibliographystyle{ieeetr}
\bibliography{refs}

\end{document}